\documentclass[prl,twocolumn,superscriptaddress]{revtex4-1}
\usepackage{graphicx}
\usepackage{dcolumn}
\usepackage{bm}
\usepackage[utf8]{inputenc}
\usepackage{float}
\usepackage{braket}
\usepackage{amsmath}

\begin{document}

\author{Peter \surname{Schattschneider}}
\email{peter.schattschneider@tuwien.ac.at}
\affiliation{Institute of Solid State Physics, TU Wien, Wiedner Hauptstra\ss e 8-10/E138-03, 1040 Wien, Austria}
\affiliation{University Service Centre for Transmission Electron Microscopy, TU Wien, Wiedner Hauptstra\ss e 8-10/E057-02, 1040 Wien, Austria}

\author{Stefan \surname{Löffler}}
\affiliation{University Service Centre for Transmission Electron Microscopy, TU Wien, Wiedner Hauptstra\ss e 8-10/E057-02, 1040 Wien, Austria}

%

\title{The electron microscope as a quantum gate}

\date{\today}

\begin{abstract}
We propose to use the topological charge instead of the spin variable to span a two-dimensional Hilbert space for beam electrons in a transmission electron microscope (TEM). In this basis, an electron can be considered as a qbit freely floating in vacuum. We show how a combination of magnetic quadrupoles with a magnetic drift tube can serve as a universal device to manipulate such qbits at the experimenter's discretion. High-end TEMs with aberration correctors, high beam coherence and utmost stability are a promising platform for such experiments,  allowing the construction of quantum logic gates   for single beam electrons in a microscope. 
\end{abstract}

\maketitle

Manipulation of the electron's phase is  an actual topic in  electron microscopy.
On the one hand, wave front engineering promises better spatial resolution, novel beam splitters~\cite{Hommelhof2015}, improved sensitivity for particular applications such as spin polarized electronic transitions~\cite{Schachinger2017},  access to otherwise undetectable physical properties such as crystal chirality~\cite{Juchtmans2015} or manipulating nanoparticles via electron vortex beams~\cite{VerbeeckAdvMat2013}.  On the other hand, the coherent control of  the  interaction of fast electrons with  electromagnetic radiation, either via near fields in PINEM~\cite{Carbone2018,Feist2015200}, resonant cavities~\cite{Baum2016} or laser accelerators~\cite{Hommelhoff2019} leads to oscillations in the probability distribution of the electron's momentum and energy, allowing the compression of fast electron pulses below the femtosecond time scale.  A similar phenomenon occurs when superpositions of Landau states propagate in a magnetic field, giving rise to rotations of  "`flower-like"' patterns~\cite{Bliokh2012}. The fascinating possibility to shape the phase of the electron wave~\cite{Verbeeck2010} with special masks or via interaction with magnetic fields, in particular with a device called mode converter that transforms a plane electron wave into one with topological charge~\cite{Schattschneider2012,Kramberger2019} led to a proposal for building  electron qbits in a traveling free electron wave~\cite{Loeffler2020}.

Here, we extend this work combining the mode converter with a magnetic drift tube.  This allows targeted manipulation of free electron  qbits, thus opening the way to construct quantum gates for logical operations.

The Schr\"odinger equation in cylindrical coordinates $(r,\phi,z)$ in a uniform magnetic field $B$ pointing in positive $z$ direction \cite{Bliokh20171}
$$
-\frac{\hbar^2}{2m}\big[\frac{1}{r} \frac{\partial}{\partial r}(r \frac{\partial}{\partial r})+\frac{1}{r^2}(\frac{\partial}{\partial \phi}+i \frac{2 r^2}{w_m^2})^2+\frac{\partial^2}{\partial z^2}\big]\Psi=E \Psi,
$$
has solutions
\begin{equation}
\psi_{k,n,l}=e^{i k z} LG_{n,l}(r, \phi)
\label{eq:1}
\end{equation}
where
$$
w_m=\sqrt{\frac{4 \hbar}{|e B|}}
$$ 
is the magnetic waist or length parameter. The pure phase factor in $z$ describes a plane wave propagating along the magnetic field direction, and the stationary solutions for the remaining $(r,\phi)$ part are Landau states~\cite{Bliokh20171}. They are recognized as non-diffracting Laguerre-Gauss modes
\begin{equation}
LG_{n,l}=(\frac{r}{w_m})^{|l|} L^{|l|}_n(\frac{2 r^2}{w_m^2})\exp(-\frac{r^2}{w_m^2}) e^{il \phi}.
\label{Landau}
\end{equation}
Here, $L^{|l|}_n$ are generalized Laguerre polynomials. The dispersion relation is~\cite{Bliokh2012}
\begin{equation}
	E_{k,n,l}=E_{k}+\hbar \Omega_L l +\hbar \Omega_L(2n+|l|+1)
	\label{eq:E}
\end{equation}
with~\footnote{In the non-relativistic case.}
$$
E_{k}=\frac{\hbar^2 k^2}{2 m}
$$
and  Larmor frequency $\Omega_L =|e B|/2 m$. 

Whereas Eq.~\ref{Landau} describes exact solutions in a magnetic field, {\it diffracting} Laguerre-Gauss functions 
\begin{eqnarray}
\chi_{k,n,l}&=&(\frac{r}{w(z)})^{|l|} L^{|l|}_n(\frac{2 r^2}{w^2(z)})\exp(-\frac{r^2}{w^2(z)}+i k\frac{r^2}{2R(z)}) \nonumber
\\
 &&e^{i(l \phi+k z}e^{i(2n+|l|+1)\zeta(z)} ,
\label{LG}
\end{eqnarray}
build an orthonormal set~\footnote{Like Bessel functions; they are orthogonal, but not normalizable.} of stationary vacuum solutions (with $B=0$) of the Schr\"odinger equation in paraxial approximation with respect to the $z$ axis.  The dispersion relation 
$$
E=\hbar^2 k^2/2 m
$$ 
defines the paraxial wave number $k$, and $w(z)=w_0\sqrt{1+z_R^2/z^2}$ is a $z$-dependent waist. $R(z)=z(1+z_R^2/z^2)$ is the wave front curvature, $\zeta(z)=\arctan(z/z_R)$ is the Gouy phase, and $z_R$ is the Rayleigh length~\footnote{These modes are diffracting, i.e. the beam waist increases for $|z|>0$.}.
For convenience we omit the energy index because $E$ is implicitly fixed to the kinetic energy of the electron beam. Comparing Eqs.~\ref{eq:1} and~\ref{Landau} with~\ref{LG}, it is evident that the two solutions can be matched at $z=0$.
The matching condition is
\begin{equation}
\chi_{k,n,l}= \psi_{k,n,l} .
\label{eq:Matching}
\end{equation}
 Eq.~\ref{eq:Matching} is fulfilled if $w_0=w_m$, i.e. if the waists of the Laguerre-Gauss mode and of the Landau mode   coincide. We are dealing with narrow wave packets of a few $\mu m$ extension, traveling in positive $z$ direction with speed $v=z/t$. Matching can be realized by switching on a homogeneous magnetic field 
\begin{equation}
B_0=
\left \{
\begin{array}{cc}
4 \hbar/|e| w_0^2 & t \geq 0 \\
0 & else
\end{array}
\right . \quad
\label{B0}
\end{equation}
 when the wave packet passes $z=0$.
We note that the above argument only holds if the magnetic field is switched on and off non-adiabatically~\cite{messiah1999quantum}. A condition for adiabaticity is that the time of change $T$ of the Hamiltonian be definitely longer than the inverse of the characteristic excitation frequency connecting adjacent energy levels~\footnote{Hommelhoff~\cite{Hommelhof2015} relies on the adiabatic principle in his proposal for a beam splitter based on a wave guide.}
$$
T>> 2 \pi\hbar/(E_n-E_{n+1})=\pi/\Omega_L=T_L/2.
$$
In other words, the magnetic field must be switched on faster than half a Larmor period $T_L$ in order that the electron does {\it not} follows adiabatically the changed eigen states. In the present example $T_L \sim 1$~ns. A $\pi$ pulse lasts $T_\pi \sim 0.5$~ns;    a long coil with low impedance could accomplish the task, having electrons fly for an odd multiple of $T_\pi$ before switching the field off. 

\begin{figure}[htbp]
	\centering
		\includegraphics[width= \columnwidth]{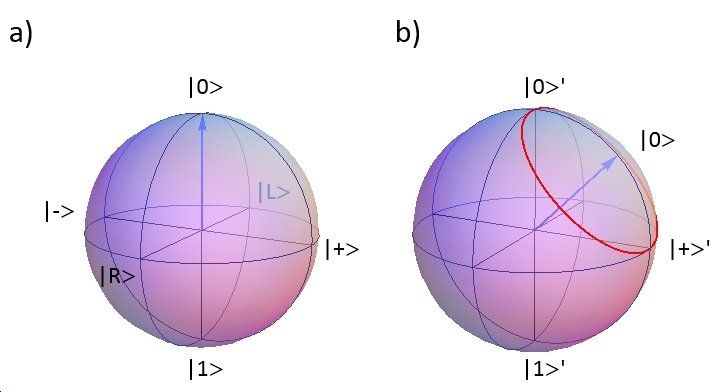}
	\caption{a) Landau states are represented by the north and south pole of the Bloch sphere. The Hermite-Gauss function, Eq.~\ref{LGx0} corresponds to the  $\ket +$ state vector. A $\pi/2$ pulse transfers it to  $\ket R $. b)Rotated coordinate system with new basis vectors $\ket 0'$ and $\ket 1 '$. The red half circle joining $\ket 0'$ an $\ket +'$  exerts the rotation corresponding to the Hadamard gate, Eq.~\ref{Hadamard}.}
	\label{fig:1}
\end{figure}

\begin{figure}[htbp]
	\centering
		\includegraphics[width= \columnwidth]{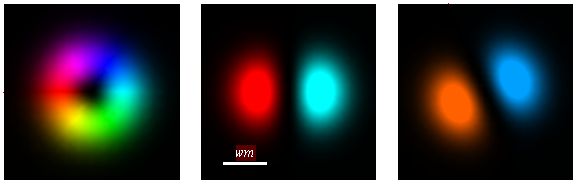}
	\caption{Qbit wave function in real space: Example of qbit manipulation with a mode converter ($\ket 0 \mapsto \ket +$) and subsequent $\pi/2$ pulse in a magnetic drift tube ($\ket + \mapsto \ket R$). The scale bar indicates one magnetic waist. Brightness codes for the probability,  color codes for the phase: The leftmost panel, corresponding to the Laguerre-Gauss mode $LG_{0,1}$, shows an azimuthal phase ramp of $2 \pi$ over the ringlike structure.}
	\label{fig:2}
\end{figure}

 One may prepare matching superpositions of  free electron states by various phase shaping methods, 
e.g.~\cite{Verbeeck2018,Vanacore2019,Shiloh2019}. In order to demonstrate the working principle we match 
\begin{equation}
\left.{ (\chi_{k,0,1} +   \chi_{k,0,-1})}\right|_{z=0} = LG_{0,1} +  LG_{0,-1}.
\label{LGx0}
\end{equation} 
 The energy of the Landau states, Eq.~\ref{eq:E} induces a time dependent phase factor on each eigen function for  $t\geq 0$: 
	\begin{equation}
	\Psi_{n,l}(t)=LG_{n,l} \, e^{i E_{n,l} t/\hbar}.
	\label{eq:hk}
	\end{equation}
Discarding a common phase factor, the wave function Eq.~\ref{LGx0}  evolves in time as
\begin{equation}
\Psi(t)=\frac{1}{\sqrt{2}} (e^{2i \Omega_L t} LG_{0,1}+LG_{0,-1}).
\label{eq:LGx1}
\end{equation}
 Eq.~\ref{eq:LGx1} describes a two state quantum system, revealing  periodic oscillations with cyclotron frequency  while the wave packet propagates along the $z$ axis.
 The state vector can thus be represented on a Bloch sphere. Using vector notation, the vector at $t=0$ in the basis of Eq.~\ref{eq:LGx1}  can be written as a qbit
$$
\frac{1}{\sqrt{2}} ( \ket{ 0} + \ket{ 1})=\ket{+}
$$
where the abbreviation $\ket 0$ corresponds to $LG_{0,-1}$ and $\ket 1$  to $LG_{0,1}$. From the definition of Hermite-Gauss functions, it follows that 
$$
\braket{r,\phi|+} =HG_{1,0}
$$ 
is a Hermite-Gauss mode.
When the wave packet propagates along $z$, this vector rotates clockwise around the $z$ axis of the Bloch sphere, (the blue vertical arrow in Fig.~\ref{fig:1}), {\it i.e.} along a latitude circle. 
As shown by L\"offler~\cite{Loeffler2020}, rotation along a meridian of this Bloch sphere can be achieved with an appropriately tuned quadrupole doublet, also known as mode converter~\cite{Schattschneider2012,Kramberger2019}. Combining the two operations, an electron qbit can be manipulated at our discretion. As an example, one could send the input qbit $\ket 0$ through a mode converter, performing $\ket 0 \mapsto \ket+$, and subsequently have it pass  a magnetic drift tube for a fraction of a Larmor period. A $\pi$ pulse corresponds to a drift time of half a Cyclotron period (or 1/4 Larmor period). It would map  $\ket + \mapsto\ket -$, and a $\pi/2$ pulse would result in   $\ket + \mapsto \ket R$, as shown in Fig.~\ref{fig:2}. 

\begin{figure}[htbp]
	\centering
		\includegraphics[width= 0.5\columnwidth]{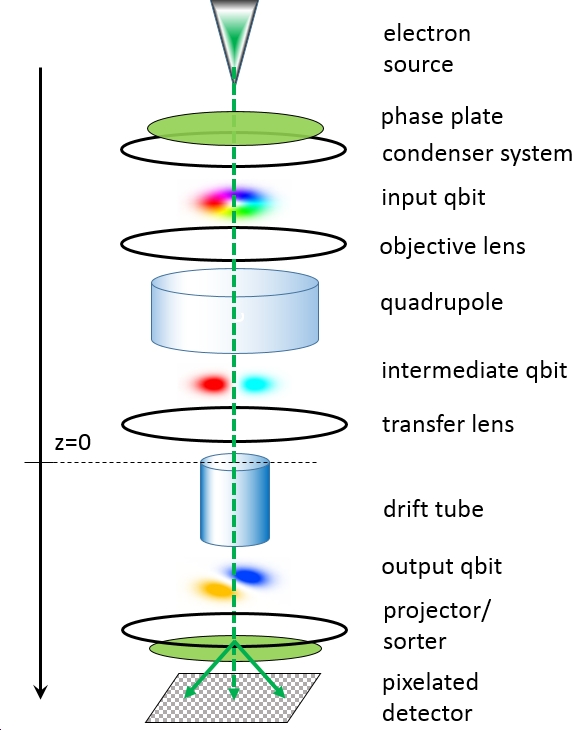}
	\caption{Column of an electron microscope with standard devices (black), phase shaping devices (green) and the main ingredients of the qbit manipulator in blue (quadrupole and drift tube). Electron qbits (color coded as in Fig.~\ref{fig:2}) travel down the positive $z$ axis. The upper rim of the drift tube is located at $z=0$. Example of the mode conversion of an input qbit  ($\ket 0 \mapsto \ket +$) and a subsequent $\pi/2$ pulse in the magnetic drift tube ($\ket + \mapsto \ket R$), as shown in Fig.\ref{fig:2}. }
	\label{fig:column}
\end{figure}

Manipulation of electron qbits on the Bloch sphere as described can be realized, provided that the beam is highly coherent, and that the parameters in Eq.~\ref{LG} can be tuned with high precision. Transmission electron microscopes (TEMs) or scanning TEMs (STEMs) are ideally suited for a proof of principle experiment, according to their versatility and performance. All necessary devices exist or can be produced with present technology. As was shown~\cite{Schattschneider2012,Kramberger2019,Loeffler2020}, the quadrupole lenses of commercial correctors can be tuned to move an input qubit along a meridian of the Bloch sphere. Phase plates or other phase shaping devices for preparing the input and projecting the output qbit onto a chosen basis~\cite{Pozzi2020} are also available. Insertion of a magnetic drift tube into the column of a (S)TEM is technically feasible. An possible experimental setup is sketched in Fig.\ref{fig:column}. Items drawn in black are standard in all TEMs. Items marked in green are non-standard, but readily available, and can easily be inserted into the instrument. Quadrupoles (light blue) are contained in any aberration corrector of high-end TEMs. The drift tube (darker blue) is the only element that needs adaptation of the instrument. The input, intermediate and output qbits sketched along the $z$-axis correspond to the example given in Fig.\ref{fig:2}.

The biggest obstacle to an experiment is probably an economic one: high-end microscopes with probe or image correctors are expensive instruments with high workload. It is rather difficult to obtain beam time for exotic experiments, and almost impossible to convince operators to open the column and insert non-standard devices.

\begin{figure}[htbp]
	\centering
		\includegraphics[width= \columnwidth]{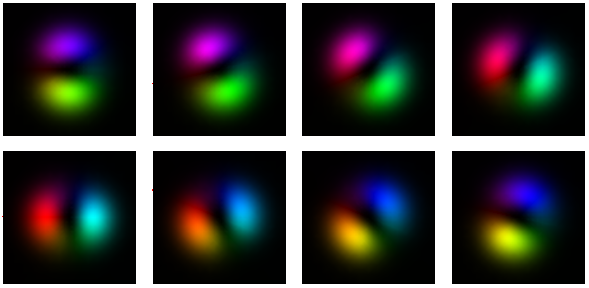}
	\caption{From left to right: Evolution of a qbit in the magnetic drift tube, starting with $\ket 0 '$, in steps of discrete drift times of 1/16 Larmor period. The panel below $\ket 0 '$ is $\ket + '$, the result of applying the Hadamard operator on the starting vector. Scale bar and color coding as in Fig.~\ref{fig:2}.}
	\label{fig:3}
\end{figure}

The manipulation of qbits is a cornerstone in quantum computing.  In order to elucidate this point further, let us discuss
the $\pi$ pulse operator mentioned above. It  rotates vectors on the Bloch sphere over the $z$ axis (represented by the blue arrow in Fig.~\ref{fig:1}) by an angle of $\pi$. Recalling the rotation operator~\footnote{
 The rotation matrices in spherical coordinates are
\begin{eqnarray*}
R_x(\theta)&=&\frac{1}{\sqrt{2}}
\begin{bmatrix}
\cos(\theta/2) & i \sin(\theta/2) \\
i \sin(\theta/2) & \cos(\theta/2)	
\end{bmatrix} 
\\
R_y(\theta)&=&\frac{1}{\sqrt{2}}
\begin{bmatrix}
\cos(\theta/2) & \sin(\theta/2) \\
-\sin(\theta/2) & \cos(\theta/2)	
\end{bmatrix} 
\\
R_z(\theta)&=&
\begin{bmatrix}
e^{i \theta/2} & 0 \\
0 & e^{-i \theta/2}	
\end{bmatrix} .
\end{eqnarray*}
}
$$
R_z(\pi)=e^{i \pi/2}
\begin{bmatrix}
	1 & 0 \cr
	0 & e^{-i \pi}
\end{bmatrix} .
$$
Apart of a global phase factor this corresponds to  the action of a Pauli-Z  gate:
$$
\hat Z :=
\begin{bmatrix}
	1 & 0 \cr
	0 & -1
\end{bmatrix} .
$$
 In passing we mention that the global phase accumulated through a closed loop on the Bloch sphere is the Berry phase~\cite{Bliokh20171,Berry1990}.
Since all states on the Bloch sphere are degenerate when the magnetic field is off, any  orthogonal basis  may be chosen. Let the transformation $\hat T$ between the old and the new basis be a rotation over the $y$-axis by $-\pi/4$. Vectors transform as:
$$
\ket{\Psi}'=\hat T \ket{\Psi}.
$$
with
$$
\hat T =R_y(\frac{\pi}{4}).
$$
With the definition of the rotation operators in the (original) basis in terms of the Pauli matrices $\sigma_{x,y,z}$
$$
R_{x,y,z}(\theta)=e^{-i \frac{\theta}{2} \sigma_{x,y,z}}
$$
 the   Pauli-Z gate in the new basis reads
\begin{equation}
\hat Z'=
\hat T \,\hat Z \, \hat T^\dag=
\frac{1}{\sqrt{2}}
\begin{bmatrix}
1& 1 \\
1 & -1	
\end{bmatrix} .
\label{Hadamard}
\end{equation}
This is the Hadamard gate, one of the better known ingredients in quantum computing. Fig.~\ref{fig:3} - from left to right - shows the evolution of a qbit along the red circle in Fig.~\ref{fig:1}, starting with $\ket 0 '$ in steps of discrete drift times of 1/16 Larmor period (or 2.75~mm steps along the $z$-axis). The panel below $\ket 0 '$ is $\ket + '$, the result of applying the Hadamard operator on the starting vector. Scale bar and color coding as in Fig.~\ref{fig:2}.
The parameters chosen for the demonstration examples of Figs.~\ref{fig:2},~\ref{fig:3}  are: $B_z=0.1$~Tesla, kinetic energy of the electron $10$~keV. This gives a Larmor frequency $\Omega_L= 8.6$~GHz, a magnetic waist $w_m=162$~nm, and an oscillation  period of $z_L=\frac{2 \pi v}{2 \Omega_L}=22$~mm along the $z$ axis. 

Another example is the rotation along a meridian $\ket 0 \mapsto \ket+$ mentioned above. The operator is
$$
\hat R_y(\pi/2)=
\begin{bmatrix}
1& -1 \\
1 & 1	
\end{bmatrix}
$$
a relative of the Hadamard operator~\footnote{Matrices with entries~$\pm 1$ are called Hadamard matrices.}. 

In conclusion, the combination of a  mode converter and a magnetic drift tube allows the design of a wide variety of quantum logic gates. High-end TEMs - instruments of utmost stability,  spatial and energy resolution, sophisticated lens systems, ultra sensitive detectors and pulsed electron sources with repetition rates of the order of GHz - provide an ideal platform to extend qbit manipulation from photons to freely floating electrons.  Recent work on entanglement in electron microscopy~\cite{Okamoto2014,Schatt2018,Schatt2019}  may even provide new possibilities to study 2-qbit gates in non-separable systems. 

Financial support by the Austrian Science Fund  under projects P29687-N36 and I4309-N36 is gratefully acknowledged.

\bibliography{Lit}

\end{document}